# A Frequency Domain Steganography using Z Transform (FDSZT)

*J. K. Mandal*

Department of Computer Science and Engineering,
University of Kalyani,
Kalyani, Nadia-741235,West Bengal, India
e-mail:- jkm.cse@gmail.com

*Abstract—Image steganography is art of hiding information onto the cover image. In this proposal a transformed domain based gray scale image authentication/data hiding technique using Z transform (ZT) termed as FDSZT, has been proposed. Z-Transform is applied on 2×2 masks of the source image in row major order to transform original sub image (cover image) block to its corresponding frequency domain. One bit of the hidden image is embedded in each mask of the source image onto the fourth LSB of transformed coefficient based on median value of the mask. A delicate handle has also been performed as post embedding operation for proper decoding. Stego sub image is obtained through a reverse transform as final step of embedding in a mask. During the process of embedding, dimension of the hidden image followed by the content of the message/hidden image are embedded. Reverse process is followed during decoding. High PSNR obtained for various images conform the quality of invisible watermark of FDSZT.*

## I. INTRODUCTION

Steganographic techniques embed secrete/authenticating information into various natural cover data like sound, images, logos etc.. Embedded data is referred to as stego-data and it must be perceptually indistinguishable from its natural cover. Steganography includes the concealment of digital information within data files/images. Generally, a steganographic message may be picture, video, sound file [6], [5]. A message may be hidden by using algorithms like invisible ink between the visible lines of innocuous documents to ensure the security which is a big concern in modern day image trafficking across the network. Security may be achieved by hiding information into images. Data hiding [4] in the image has become an important tool for image authentication. Ownership verification and authentication are the major task for military people, research institute and scientists. Information security and image authentication has become very important to protect digital image document from unauthorized access [2]. Data hiding refers to the nearly invisible [3], [7], [12], [13], [14] embedding of information within a host data set as message, image or video. A classic example of steganography is that of a prisoner communicating with the outside world under the supervision of a warden. The data hiding represents a useful alternative to the construction of a hypermedia document or image, which is very less convenient to manipulate. The goal of steganography is to hide the message/image in the source image by some key techniques and cryptography is a process to hide the message content. The motive is to hide a message inside an image keeping its visible properties [8] the source image as close to the original. The most common methods to make these alteration is usage of the least-significant bit (LSB) developed through[8] masking, filtering and transformations on the source image[15]. Present proposal is an algorithm ich would facilitate secure message transmission through block based data hiding. Most of the works [11],[10], [1] used minimum bits of the hidden image for embedding in spatial domain, but the proposed algorithm embed in transformed domain with a bare minimum distortion of visual property.

Rest of the paper is organized as follows. Section II deals with the proposed technique. Results and comparisons are given in section III. Concluding remarks are presented in section IV and references are drawn at end.

## II. THE TECHNIQUE

In the process of embedding a 2 x 2 mask is chosen in row major order. One bit of the authenticating message/image is embedded in each mask rowwise in transformed domain. 2×2 gray scale image mask is transformed from spatial domain to frequency domain using Z-Transform. The dimension of the hidden image is extracted. Along with the hidden image, the dimensional values are also embedded into the real part of the host image mask on fourth LSB bit of the transformed coefficient within 2×2 mask, where the coefficient is chosen based on median lue of the coefficient of 2 x 2 mask. Embedded mask is transformed from frequency domain to spatial domain using inverse Z-Transform. The formula for Z- Transform is

$$X(z) = \sum_{m=0}^{\alpha} x(m)\, r^{-m}\, e^{-j\omega n}$$ (limit is taken 0 to α as pixel value cannot be negative for an image)

In the present implementation the value of r is taken as 1 and ω varies between $0 \leq \omega \leq 2\pi$. For a 2×2 sub image there are four pixel values in the mask and set of frequencies taken are: $\omega = \{ 0, \pi/2, \pi, 3\pi/2 \}$.

Schematic diagram of the technique is shown Figure1.

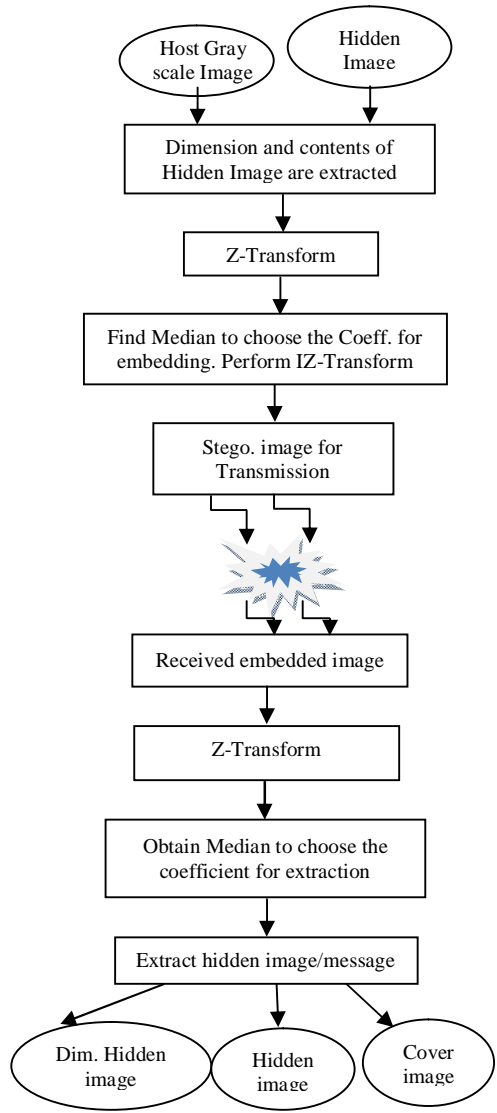

Figure 1: Schematic diagram of FDSZT

Algorithm of insertion and extraction are given in section A and B respectively. A complete example has also been illustrated in section C.

### A. Algorithm for insertion

The technique uses gray scale image of size p×q as input. Hidden image of size m×n is chosen. One bit of hidden image is embedded in each mask based on median values of transformed coefficients in Z-domain. Z-Transform is a two dimensional function where (n1, n2) is a spatial coordinate can be represented as

$$f(z1, z2) = \sum_{n1=-\infty}^{\infty} \sum_{n2=-\infty}^{\infty} f(n1, n2) z1^{-n1} z2^{-n2} \quad (1)$$

where z1 and z2 are both complex numbers consisting of real and an imaginary parts. Since z1 and z2 are complex numbers, let $z1=e^{j\omega 1\pi}$ and $z2=e^{j\omega 2\pi}$, Where $e^{j\theta} = \cos\theta + j\sin\theta$. Substituting the values of z1 and z2 in equation (1), the equation becomes the discrete form of two dimensional Z-Transformation equation.

$$f(e^{j\omega 1\pi}, e^{j\omega 2\pi}) = \sum_{n1=-\infty}^{\infty} \sum_{n2=-\infty}^{\infty} f(n1, n2) e^{j\omega 1\pi^{-n1}} e^{j\omega 2\pi^{-n2}}$$

$$\text{Or } f(\omega 1, \omega 2) = \sum_{n1=-\infty}^{\infty} \sum_{n2=-\infty}^{\infty} f(n1, n2) e^{-j\pi(n1\omega 1 + n2\omega 2)} \quad (2)$$

where $\omega_1$ and $\omega_2$ are two frequency variables, varies from -∞ to + ∞ and n1 and n2 is finite and positive numbers. In case of present implementation ω ranges between 0 to3π/2.

**Algorithm:**

Input: Host image of size p×q, hidden image of size m×n.
Output: Embedded image of size p×q.
Method: Insertion of hidden image bitwise into the gray scale image.

*Step 1:* Obtain the size of the hidden image m×n
*Step 2:* For each hidden message/image, read source image mask of size 2×2 in row major order. Apply Z-Transform onto the selected cover image mask(2 x 2) to obtain coefficients in transformed domain.
*Step 3:* Obtain Median of the four frequency coefficients obtained in step 2 to choose the byte for embedding.
*Step 4:* Embed 1 secret bit onto the fourth LSB position towards left of the byte.
*Step 5:* Apply adjustmet if necessary so that the coefficient will be the median of the mask after embedding
*Step 5:* Apply IZ-Transform to back the mask from Z domain to spatial domain.
*Step 6:* Repeat step 2 to 6 for the whole cover image.
*Step 7:* Stop

### B. Algorithm for extraction

The hidden image is received in spatial domain. The embedded image is taken as the input and the hidden message/ image size, content are extracted from it in transform domain. The continuous Inverse Z-Transform of a function f(n1, n2) is represented as

$$f(n1,n2) = \left(\frac{1}{2\pi j}\right)^2 \oiint f(z1,z2) z1^{n1-1} z2^{n2-1} dz1\, dz2 \quad (3)$$

where $f(n1,n2)$ be a function and $f(z1,z2)$ be the Z-Transform of the function $f(n1,n2)$. Control integration is for irregular spaces in z-domain. Since $z1$ and $z2$ are complex numbers, Let $z1=e^{j\omega 1\pi}$ and $z2=e^{j\omega 2\pi}$, where $e^{j\omega\theta}= \cos\omega\theta + j\sin\omega\theta$. Substituting the values of $z1$ and $z2$ in equation (3), we have a discrete form of inverse Z Transform for two dimensions. Now $z1=e^{j\omega 1\pi}$, differentiating this with respect to we get $\frac{dz1}{d\omega 1} = e^{j\omega\omega 1\pi} j\pi$, therefore $dz1=e^{j\omega 1\pi} j\pi\, d\omega 1$ and $z2=e^{j\omega 2\pi}$, differentiating this with respect to $\omega 2$ we get $\frac{dz2}{d\omega 2} = e^{j\omega\omega 2\pi} j\pi$, therefore $dz2=e^{j\omega 2\pi} j\pi\, d\omega 2$. The equation (3) becomes from the above derivation is

$$f(n1,n2) = \left(\frac{1}{2\pi j}\right)^2 \oiint f(e^{j\omega 1\pi}, e^{j\omega 2\pi}) e^{j\omega 1\pi^{n1-1}} e^{j\omega 2\pi^{n2-1}} e^{j\omega 1\pi} j\pi\, d\omega 1\, e^{j\omega 2\pi} j\pi\, d\omega 2$$

The discrete form of this control integration equation is as follows

$$f(n1,n2) = \frac{1}{4} \sum_{\omega 1=-1}^{1} \sum_{\omega 2=-1}^{1} f(\omega 1,\omega 2) e^{j\pi(n1\omega 1+n2\omega 2)} \quad (4)$$

The equation (4) is the discrete form of Two Dimensional Inverse Z Transform

**Algorithm:**
  Input: Embedded image of size p×q.
  Output: Host image of size p×q, hidden image of size m×n.
  Method: Extract bits of hidden image from embedded image
  *Step 1:* Read embedded image mask( of size 2×2) in row major order. Apply Z-Transform onto the embedded image mask to transform the embedded sub image from spatial to frequency domain so that four frequency components are regenerated.
  *Step 2:* Obtain Median of four frequency components to choose the embedded byte from 2×2 mask.
  *Step 3:* Extract the secret bit from the byte embedded in fourth LSB position. Replace hidden message/ image bit position in the block by '1'. For each eight extracted bits construct one image pixel of authenticating image.
  *Step 4:* Repeat step 1 to 3 to regenerate hidden image as per size of the hidden image.
  *Step 5:* Stop

*C. Example*

Consider a byte of Jet image (figure 2a ) to be inserted into each mask of Lenna image (Figure 2c). Figure 2b shows pixels of Lenna image in spatial domain. One bit of the Jet image is inserted into byte of the Lenna image in 2×2 mask. Insertion is done in the rightmost fourth LSB bit of the byte of Lenna. Resultant image after embedding is shown in Figure 2d in frequency domain and Figure 2e in spatial domain.

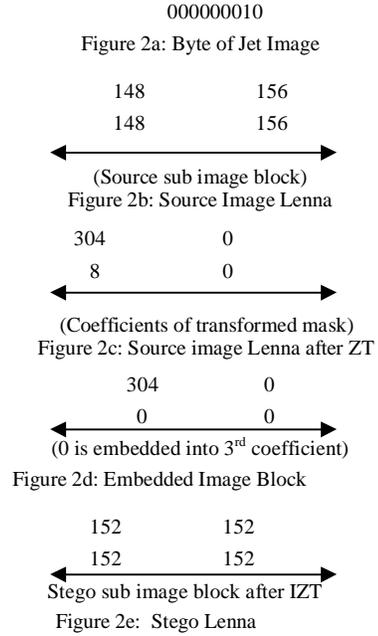

000000010
Figure 2a: Byte of Jet Image

| 148 | 156 |
| 148 | 156 |

(Source sub image block)
Figure 2b: Source Image Lenna

| 304 | 0 |
| 8 | 0 |

(Coefficients of transformed mask)
Figure 2c: Source image Lenna after ZT

| 304 | 0 |
| 0 | 0 |

(0 is embedded into 3rd coefficient)
Figure 2d: Embedded Image Block

| 152 | 152 |
| 152 | 152 |

Stego sub image block after IZT
Figure 2e: Stego Lenna

Figure 2. Encoding process of FDSZT

## III. RESULTS, COMPARISON AND ANALYSIS

Extensive analysis has been made on various images using FDSZT technique. This section represents the results, discussions in terms of visual interpretation and peak signal to noise ratio. Figure 3a shows the host images Lenna, Tiffany, Cameraman. Figure 3b shows embedded Lenna, Tiffany, Cameraman on embedding Jet image using FDSZT. Figure 3c is the authenticating image Jet. Table I show the PSNR values for each embedding against the source image. From the table it is seen that the maximum value of the PSNR is 43.100029 and that of minimum value of the PSNR is 40.824833. The value of the PSNR is consistent for various images. The following formula is used to calculate PSNR, MSE and IF(image fidelity).

$$MSE = \frac{1}{MN} * \sum_{m,n} (I_1\, m,n - I_2\, m,n)^2$$
$$PSNR = 10\, \log(max(I_{m,n}^2)/MSE)$$

$$IF = 1 - \sum_{m,n}(I_{1_{m,n}} - I_{2_{m,n}})^2 / \sum_{m,n} I_{2_{m,n}}^2$$

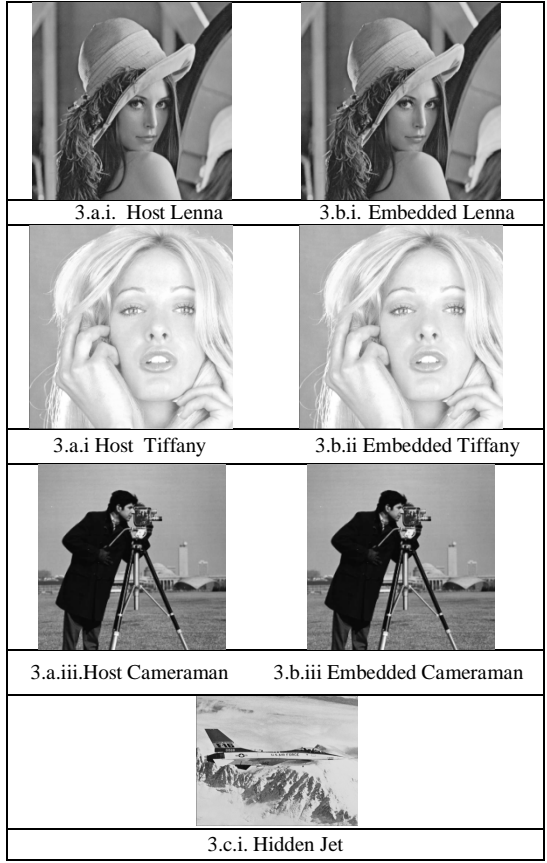

3.a.i. Host Lenna     3.b.i. Embedded Lenna

3.a.i Host Tiffany     3.b.ii Embedded Tiffany

3.a.iii.Host Cameraman     3.b.iii Embedded Cameraman

3.c.i. Hidden Jet

Figure 3: Visual effect of embedding in FDSZT

TABLE I.    PSNR, MSE, IF VALUES OBTAINED FOR VARIOUS IMAGES USING FDSZT

| *Host Image* | *PSNR values* | *MSE Values* | *IF* |
|---|---|---|---|
| Lenna | 41.620922 | 4.477013 | 0.999722 |
| Tiffany | 40.824833 | 5.377705 | 0.999882 |
| Cameraman | 43.100029 | 3.184765 | 0.999823 |

## IV. CONCLUSION

The proposal is a novel embedding approach termed as, FDSZT based on Z Transformation for gray scale images where concept of median has been used to select the coefficient for embedding in Z-Transformed domain. From experimental results it is clear that the proposed technique obtained consistent PSNR ratio along with good image fidelity for various images which conform that Z-transformed based image steganography can obtain better visibility/quality. Payload may be increased considerably which is the future scope of the paper and hence research in Z-Domain.